\documentclass[a4paper]{jpconf}
\usepackage{graphicx}

\usepackage{cite}
\usepackage{amsbsy}
\def\be{\begin{equation}}
\def\ee{\end{equation}}
\def\beqn{\begin{eqnarray}}
\def\eeqn{\end{eqnarray}}
\def\no{\nonumber}
\def\ba{\begin{array}{c}}
\def\bat{\begin{array}{cc}}
\def\ea{\end{array}}
\def\bi{\begin{itemize}}
\def\ei{\end{itemize}}

\def\cL{{\cal L}}

\def\cO{{\cal O}}

\newcommand{\eqn}[1]{(\ref{#1})}
\newcommand{\bel}[1]{\be\label{#1}}
\newcommand{\rms}{\rm\scriptsize}

\begin{document}

\title{Theoretical overview of kaon decays}

\author{Antonio Pich}

\address{IFIC, Universitat de Val\`encia -- CSIC, Apt. Correus 22085, E-46071 Val\`encia, Spain}

\ead{Antonio.Pich@ific.uv.es}

\begin{abstract}
Kaon decays are an important testing ground of the electroweak flavour theory. They can provide new signals of CP violation and, perhaps, a window into physics beyond the Standard Model. At the same time, they exhibit an interesting interplay of long-distance QCD effects in flavour-changing transitions.
A brief overview is presented, focusing on a few selected topics of particular interest.
A more detailed and comprehensive review can be found in Ref.~\cite{Cirigliano:2011ny}.
\end{abstract}

\section{Effective Field Theory}  

Kaon physics has been at the origin of many fundamental ingredients of the Standard Model (SM), such as flavour quantum numbers, parity violation, meson-antimeson mixing, quark mixing, CP violation and the GIM mechanism~\cite{Cirigliano:2011ny}. Rare kaon decays provide sensitivity to short-distance scales ($c$, $t$, $W^\pm$, $Z$) and offer the exciting possibility of unravelling new physics beyond the SM. Searching for forbidden lepton-flavour-violating processes
($K_L\to e^\pm \mu^\mp$, $K_L\to e^\pm e^\pm \mu^\mp\mu^\mp$, $K^+\to\pi^+\mu^\pm e^\mp$\ldots)
beyond the $10^{-10}$ level, one is actually exploring energy scales above the 10 TeV region.
The study of allowed decay modes provides, at the same time, very interesting tests of the SM
itself, improving our understanding of the interplay among electromagnetic, weak and strong interactions. In addition, new signals of CP violation, which would help to elucidate the source of CP-violating phenomena, can be looked for.

Owing to the presence of very different mass scales ($m_\pi < m_K \ll M_W$), the QCD corrections are amplified by large logarithms. The short-distance logarithmic corrections can be summed up, using the operator product expansion (OPE) and the renormalization group, all the way down from $M_W$ to scales $\mu < m_c$ \cite{Gilman:1979bc}. One gets in this way an  effective Lagrangian, defined in the three-flavour theory, which is a sum of local four-fermion operators $Q_i$,
constructed with the light degrees of freedom ($u,d,s; e,\mu,\nu_\ell$), modulated by
Wilson coefficients $C_i(\mu)$ which are functions of the heavy ($Z,W,t,b,c,\tau$) masses:
\bel{eq:sd_hamiltonian}
\cL_{\mbox{\rms eff}}^{\Delta S=1} \; = \; -{G_F\over\sqrt{2}}\,
V_{ud}^{\phantom{*}} V_{us}^*\;
\sum_i\, C_i(\mu)\, Q_i \, , 
\qquad\qquad
C_i(\mu)\; =\; z_i(\mu) -  y_i(\mu)\, \frac{V_{td}^{\phantom{*}}V_{ts}^*}{V_{ud}^{\phantom{*}} V_{us}^*}\, .
\ee
The CP-violating decay amplitudes are proportional to the components $y_i(\mu)$.
The overall renormalization scale $\mu$
separates the short- ($M>\mu$) and long-distance ($m<\mu$)
contributions,  which are contained in $C_i(\mu)$
and $Q_i$, respectively.
The Wilson coefficients are fully known at the next-to-leading order (NLO)
\cite{Buras:1992tc,Ciuchini:1995cd};
this includes all corrections of $\cO(\alpha_s^n t^n)$ and $\cO(\alpha_s^{n+1} t^n)$,
where $t\equiv\log{(M_1/M_2)}$ refers to the logarithm of any ratio of
heavy mass scales ($M_{1,2}\geq\mu$).
In order to calculate the kaon decay amplitudes,
we also need to know the non-perturbative matrix elements of the operators
$Q_i$ between the initial and final hadronic states.

\begin{figure}[t]
\begin{minipage}[c]{.5\linewidth}\centering
\setlength{\unitlength}{0.46mm}          
\begin{picture}(163,133)
\put(0,0){\makebox(163,133){}}
\thicklines
\put(8,124){\makebox(25,10){Energy}}
\put(43,124){\makebox(42,10){Fields}}
\put(101,124){\makebox(52,10){Effective Theory}}
\put(5,123){\line(1,0){153}} {
\put(8,86){\makebox(25,30){$M_W$}}
\put(43,86){\framebox(42,30){\fontsize{10}{12}\selectfont $\ba W, Z, \gamma, g \\
     \tau, \mu, e, \nu_i \\ t, b, c, s, d, u \ea $}}
\put(101,86){\makebox(52,30){Standard Model}}

\put(8,43){\makebox(25,20){$\lsim m_c$}}
\put(43,43){\framebox(42,20){\fontsize{10}{12}\selectfont $\ba  \gamma, g  \, ;\, \mu ,  e, \nu_i \\ s, d, u \ea $}}
\put(101,43){\makebox(52,20){$\cL_{\mathrm{QCD}}^{N_f=3}$,
             $\cL_{\mathrm{eff}}^{\Delta S=1,2}$}}

\put(8,0){\makebox(25,20){$m_K$}}
\put(43,0){\framebox(42,20){\fontsize{10}{12}\selectfont $\ba\gamma \; ;\; \mu , e, \nu_i  \\
            \pi, K,\eta  \ea $}}
\put(101,0){\makebox(52,20){$\chi$PT}}
\linethickness{0.3mm}
\put(64,39){\vector(0,-1){15}}
\put(64,82){\vector(0,-1){15}}
\put(69,72){OPE}
\put(69,29){$N_C\to\infty $}}                     
\end{picture}
\vskip -.3cm\mbox{}
\caption{Evolution from $M_W$ to $m_K$. 
  \label{fig:eff_th}}
\end{minipage}
\hfill
\begin{minipage}[c]{.43\linewidth}\centering
\mbox{}\vskip -.5cm
\includegraphics[width=6.2cm]{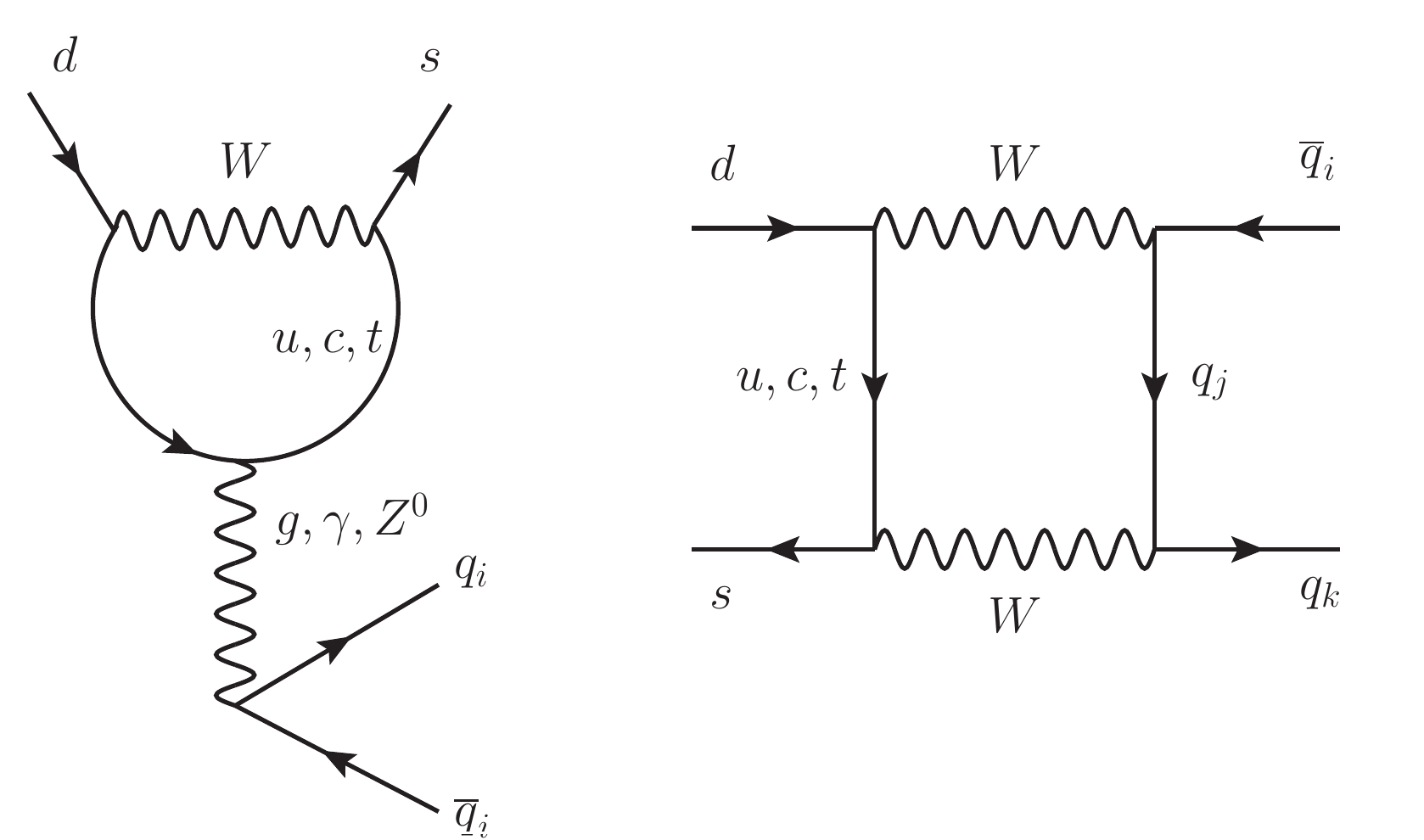}
\caption{\label{fig:sd-diagrams}
Short-distance diagrams.} 
\vskip .3cm
\includegraphics[width=6.5cm]{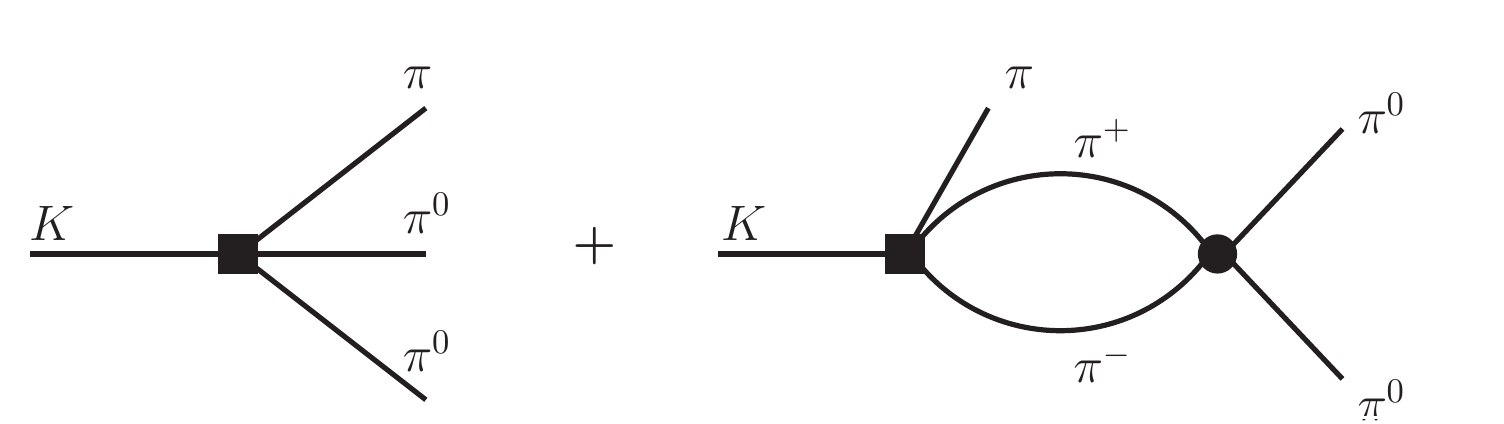}
\caption{\label{fig:sd-diagrams}
Long-distance diagrams.}
\end{minipage}
\end{figure}

At low energies, one can use symmetry considerations to define another effective field theory in terms of the QCD Goldstone bosons ($\pi, K, \eta$). Chiral Perturbation Theory ($\chi$PT) \cite{Weinberg:1978kz,Gasser:1984gg}
describes the pseudoscalar-octet dynamics through a perturbative expansion in powers of momenta and quark masses over the chiral symmetry-breaking scale $\Lambda_\chi\sim 1$~GeV.
Chiral symmetry fixes the allowed operators, while all short-distance information is encoded in their low-energy couplings (LECs) \cite{Ecker:1994gg,Pich:1995bw}.
At LO the most general effective Lagrangian, with the same
$SU(3)_L\otimes SU(3)_R$ transformation properties as the short-distance
Lagrangian \eqn{eq:sd_hamiltonian}, contains three terms \cite{Cirigliano:2011ny}:
\bel{eq:lg8_g27}
\cL_2^{\Delta S=1} = -{G_F \over \sqrt{2}}  V_{ud}^{\phantom{*}} V_{us}^*
\left\{ g_8  \,\langle\lambda L_{\mu} L^{\mu}\rangle   +
g_{27} \left( L_{\mu 23} L^\mu_{11} + {2\over 3} L_{\mu 21} L^\mu_{13}\right) +
e^2 g_8  g_{\rms ew} F^6\, \langle\lambda U^\dagger Q U\rangle
\right\} ,
\ee
where $U\equiv \exp(i \vec{\lambda} \vec{\phi} /F)$ parameterizes the Goldstone fields,
$L_{\mu}=i F^2 U^\dagger D_\mu U$  represents the octet of
$V-A$ currents, $\lambda\equiv (\lambda_6 - i \lambda_7)/2$ projects onto the
$\bar s\to \bar d$ transition, 
$Q={1\over 3} \,\mbox{\rm diag}(2,-1,-1)$ is the quark charge matrix
and $\langle\, \rangle$ denotes a 3-dimensional flavour trace.
The LECs $g_8$ and $g_{27}$ measure the strength of the two
parts of $\cL_{\mbox{\rms eff}}^{\Delta S=1}$  
transforming as
$(8_L,1_R)$ and $(27_L,1_R)$, respectively, under chiral rotations, while $g_{\rms ew}$
accounts for the electromagnetic penguin operators.

The $\chi$PT framework determines the most general form of the
K decay amplitudes, compatible with chiral symmetry,
in terms of the LECs multiplying the relevant chiral operators.
These LECs, which encode the short-distance dynamics,
can be determined phenomenologically and/or calculated
in the limit of a large number of QCD colours $N_C$. 
Chiral loops generate non-polynomial contributions, with
logarithms and threshold factors as required by unitarity.
Fig.~\ref{fig:eff_th} shows schematically the procedure used
to evolve down from $M_W$ to $m_K$.
While the OPE resums the short-distance logarithmic corrections
$\log{(M/\mu)}$, the $\chi$PT loops take care of the large
infrared logarithms  $\log{(\mu/m_\pi)}$ associated with unitarity corrections
(final-state interactions).

\section{Leptonic and Semileptonic Decays}

In (semi)leptonic decays
strong interactions only appear through the hadronic matrix elements of the left-handed 
current, which can be precisely studied within $\chi$PT and with lattice simulations.

The ratio $\Gamma[K^-\to e^-\bar\nu_e (\gamma)]/\Gamma[K^-\to \mu^-\bar\nu_\mu (\gamma)]$
has been calculated \cite{Marciano:1993sh,Cirigliano:2007xi}
and measured \cite{Lazzeroni:2012cx,Ambrosino:2009aa}
with high accuracy, allowing for a precise test of charged-current lepton universality: $|g_\mu/g_e| = 0.9978\, (20)$.
Similar precisions have been achieved in $K\to\pi\ell\nu_\ell$ \cite{Antonelli:2010yf},
$\pi\to \ell\nu_\ell$
and $\tau\to\nu_\tau\ell\nu_\ell$ \cite{Pich:2013lsa} decays.
The ratio 
$\Gamma[K^-\to \mu^-\bar\nu_\mu (\gamma)]/\Gamma[\pi^-\to \mu^-\bar\nu_\mu (\gamma)]$
provides information on the quark mixing matrix \cite{Antonelli:2010yf,Marciano:2004uf}.
With a careful treatment of electromagnetic and isospin-violating corrections,
one extracts $|V_{us}/V_{ud}| |F_K/F_\pi| = 0.2763 \pm 0.0005$ \cite{Cirigliano:2011tm}. Taking for the ratio of 
decay constants the ($N_f=2+1+1$) lattice average $F_K/F_\pi = 1.196 \pm 0.005$ \cite{Aoki:2013ldr}, this gives
\be\label{eq:Vus_ud}
|V_{us}/V_{ud}| \; =\; 0.2310 \pm 0.0011\, .
\ee

Including electromagnetic and isospin-breaking corrections \cite{Cirigliano:2008wn,Kastner:2008ch},
the most recent $K_{\ell 3}$ data leads to
$|V_{us}\, f_+(0)| = 0.2163\pm 0.0005$ \cite{Antonelli:2010yf},
with $f_+(0)= 1 + \cO[(m_s-m_u)^2]$ the $K^0\to\pi^-\ell^+\nu_\ell$ vector form factor.
The exact value of $f_+(0)$ has been thoroughly investigated since the first
precise estimate by Leutwyler and Roos, $f_+(0) = 0.961\pm 0.008$ \cite{Leutwyler:1984je}.
While analytical calculations based on $\chi$PT obtain
higher values \cite{Jamin:2004re,Cirigliano:2005xn}, owing to the large 
($\sim +0.01$) 2-loop corrections \cite{Bijnens:2003uy},
lattice results \cite{Aoki:2013ldr} used to agree with the Leutwyler--Roos estimate.
The most recent lattice analyses \cite{Bazavov:2013maa,Boyle:2013gsa}
give, however, larger values in better agreement with the $\chi$PT expectations.
The FLAG $N_f=2+1$ average $f_+(0) = 0.9661\, (32)$ \cite{Aoki:2013ldr} implies
$|V_{us}| = 0.2239\, (9)$, while a recent $N_f=2+1+1$ determination \cite{Bazavov:2013maa},
$f_+(0) = 0.9704\, (32)$, results in
\bel{eq:Vus}
|V_{us}|\, =\, 0.2229\pm 0.0005_{\rms exp}\pm 0.0007_{\rms th} \, .
\ee
Together with $|V_{ud}| = 0.97425 \pm 0.00022$
and the tiny $|V_{ub}|$ contribution \cite{Agashe:2014kda}, Eqs.~\eqn{eq:Vus_ud} and \eqn{eq:Vus} imply an stringent test of the unitarity of the
quark mixing matrix:
\be
\Delta_{\rms CKM} = |V_{ud}|^2  +  |V_{us}|^2  +  |V_{ub}|^2  - 1 \; =\;  -0.0007 \pm 0.0007\, .
\ee

\section{Non-leptonic Decays}

The measured $A(K\to\pi\pi)_I$ decay amplitudes show a strong enhancement of the octet
$\Delta I =\frac{1}{2}$ transition amplitude into a $2\pi$ final state with isospin $I=0$:
$|A_0/A_2|\approx 22$. In the $\chi$PT framework
this manifests as a huge difference between the LECs in Eq.~\eqn{eq:lg8_g27}. A LO fit to
the data gives
$\vert g_8 \vert \simeq 5.0$ and $\vert g_{27} \vert \simeq 0.285$.
Part of the enhancement originates in the strong rescattering of the final pions, which at one loop increases $A_0$
by roughly 35\%. Taking the $\chi$PT 1-loop contributions into account, one finds a sizeably smaller octet coupling
(the central values change slightly to 3.61 and 0.297, respectively, if isospin violation is included)
\cite{Cirigliano:2011ny,Cirigliano:2009rr,Cirigliano:2003gt}:
\bel{eq:g8-g27}
\vert g_8 \vert\, =\, 3.62\pm 0.28\, ,\qquad \qquad\qquad
\vert g_{27} \vert \, =\, 0.286\pm 0.028\, .
\ee
In the absence of QCD, the SM ($W$ exchange) prediction $g_8 = g_{27} = \frac{3}{5}$
would disagree with \eqn{eq:g8-g27}. The short-distance QCD corrections show the needed qualitative trend to understand the data. The matching of the effective descriptions $\cL_{\mbox{\rms eff}}^{\Delta S=1}$ (short-distance) and
$\cL_2^{\Delta S=1}$ ($\chi$PT) can be done in the large-$N_C$ limit;
including the large short-distance logarithmic corrections $\sim \frac{1}{N_C}\,\log{(M_W/\mu)}$, this gives the results
$g_8^\infty = 1.13 \pm 0.18$ and $g_{27}^\infty = 0.46 \pm 0.01$ \cite{Cirigliano:2011ny}, which show the relevance of missing NLO corrections in $1/N_C$.

\begin{figure}[t]
\begin{minipage}[c]{.4\linewidth}\centering
\includegraphics[width=5.2cm]{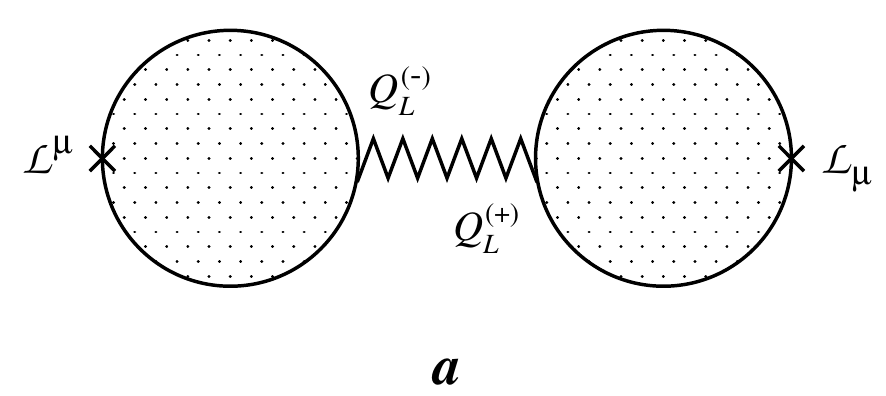}
\end{minipage}
\hfill
\begin{minipage}[c]{.56\linewidth}\centering
\includegraphics[width=7.7cm]{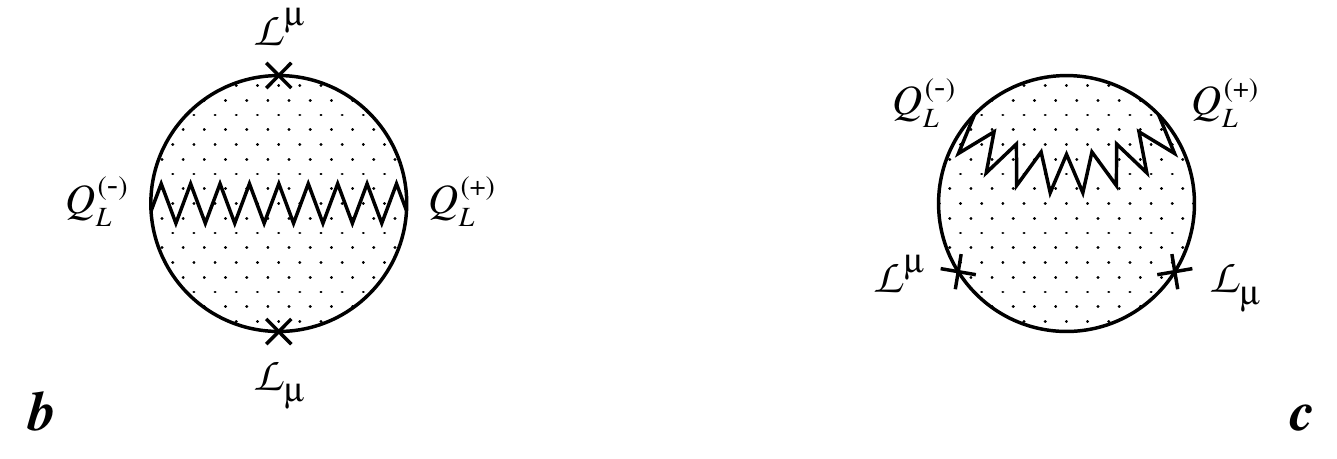}
\end{minipage}
\caption{\label{fig:K2pTopologies}
$K\to\pi\pi$ topologies. The solid lines represent quark fields propagating in a gluon background simulated by the dotted lines. The $W^\pm$ is exchanged between two $Q_L^{(\pm)}$ sources~\cite{Pich:1995qp}.}
\end{figure}

A dynamical understanding of the $\Delta I=\frac{1}{2}$ enhancement was achieved long time ago \cite{Pich:1995qp}, through a combined expansion in powers of momenta ($\chi$PT) and $1/N_C$.
With one virtual $W^\pm$ field emitted and reabsorbed, and to LO in the chiral expansion
(two $L_{\mu}$ insertions at most), there are three possible chiral invariant configurations which give rise to the effective Lagrangian
\bel{eq:EffCoup}
\cL_{\mathrm{eff}} \, = \, -\frac{G_F}{\sqrt {2}}\,
\left\{ a\;\langle Q_{L}^{^{(-)}} L_{\mu}\rangle\, \langle Q_{L}^{^{(+)}} L^{\mu}\rangle
\, +\, b\; \langle Q_{L}^{^{(-)}} L_{\mu} Q_{L}^{^{(+)}} L^{\mu}\rangle
\, +\, c\;\langle Q_{L}^{^{(-)}} Q_{L}^{^{(+)}} L_{\mu} L^{\mu}\rangle\right\}\, .
\ee
The operators $Q_L^{(+)} = Q_L^{(-)\dagger}$ represent the emission and absorption of the
virtual $W^\pm$ field; thus,
$[Q_L^{(+)}]_{ij} = \delta_{i1}\delta_{j2} V_{ud} + \delta_{i1}\delta_{j3} V_{us}$.
The underlying functional integral over quark and gluon fields which gives rise to the effective couplings in \eqn{eq:EffCoup} is represented diagrammatically in Fig.~\ref{fig:K2pTopologies}.
When further restricted to $\Delta S = 1$ transitions, the effective Lagrangian \eqn{eq:EffCoup} reduces to \eqn{eq:lg8_g27} with \cite{Pich:1995qp}
\bel{eq:g_res}
g_8\, =\,\frac{3}{5}\, (a+b)-b+c\, ,
\qquad\qquad\qquad
g_{27}\, =\,\frac{3}{5}\, (a+b)\, .
\ee

The topology which leads to the $a$-type coupling is $\cO(N_C^2)$, while those generating $b$ and $c$ are $\cO(N_C)$.
In the large-$N_C$ limit the coupling $a$ can be calculated because the four-quark operators factorize into QCD currents with well-known $\chi$PT realizations.
This factorization is only broken by (at least two) gluonic exchanges which are of NNLO in the $1/N_C$ expansion. The $c$-type configuration corresponds to the so-called penguin-like diagrams which can also be calculated at LO in terms of known phenomenological parameters. Taking into account the factor $F^2\sim \cO(N_C)$, included in the definition of the currents $L_\mu$, one easily finds \cite{Pich:1995qp}:
\beqn\label{eq:a-b}
a & = & 1 \, +\,\cO ( 1/N_C^2)\, ,
\qquad\qquad\qquad\qquad
b \; = \; \cO ( 1/N_C)\, ,
\no\\
c & =&\mathrm{Re}\, C_{4}-16\,L_{5}\,\mathrm{Re}\, C_{6}(\mu)\,
\left( \langle 0|\bar{q} q |0\rangle (\mu)/F^3 \right)^2
+\,\cO ( 1/N_C^2)\;\simeq\; 0.3\pm 0.2\, ,
\eeqn
where $L_5$ is an $\cO(p^4)$ coupling of the strong $\chi$PT Lagrangian. To $\cO (1/N_C)$ the scale dependence in $C_6(\mu)$ cancels with the one in the quark condensate
$\langle 0|\bar{q} q |0\rangle (\mu)$, while $C_4$ is scale-independent.

The non-leading topology $b$ cannot be evaluated in a model-independent way.
The important observation made in Ref.~\cite{Pich:1995qp} is that it contributes with opposite signs to $g_8$ and $g_{27}$. Taking Eq.~\eqn{eq:a-b} into account, the experimental value of $A_2$, {\it i.e.} $\vert g_{27} \vert \approx 0.29$, implies~\cite{Pich:1995qp}:
\bel{eq:b-result}
b \approx -0.52  + \cO (1/N_C^2)\, ,
\qquad\qquad\qquad
g_8\, \approx\, 1.1 + \cO (1/N_C^2)\, .
\ee
Thus, the measured ratio $\vert g_8/g_{27}\vert $ requires a large and negative value of $b$, generating a significant cancellation in $g_{27}$ and a sizeable enhancement of $g_8$.
This is precisely what was previously predicted through model-dependent calculations \cite{Pich:1990mw,Bertolini:1997ir,Bardeen:1986vz,Bijnens:1998ee} and confirmed through a rigorous inclusive NLO analysis of the two-point correlators associated with the short-distance Lagrangians $\cL_{\mbox{\rms eff}}^{\Delta S=1,2}$ \cite{Pich:1995qp,Pich:1990mw,Jamin:1994sv}.
Recently, the predicted cancellation in $g_{27}$ has been observed by a lattice calculation of $A_2$ which finds $b/a \approx -0.7$ \cite{Boyle:2012ys}; the corresponding enhancement of $g_8$ is also seen in $A_0$, although the present lattice results are still obtained at unphysical kinematics.

A similar cancellation is observed in lattice simulations of the $\Delta S = 2$ transition amplitude \cite{Lellouch:2011qw,Carrasco:2013jda}. In the chiral limit, the so-called $B_K$ parameter which regulates the $K^0$--$\bar K^0$ hadronic matrix element is given by $B_K = \frac{3}{4} (a+b)$~\cite{Pich:1995qp}, which behaves as $g_{27}$. This result agrees
with explicit model calculations \cite{Pich:1990mw,Bertolini:1997ir,Bardeen:1986vz,Bijnens:1998ee,Bijnens:2006mr,Peris:2000sw}
and QCD sum-rule determinations \cite{Pich:1985ab,Prades:1991sa}.

\section{Direct CP Violation: $\boldsymbol{\varepsilon'/\varepsilon}$}

The measured CP-violating
ratio \cite{Batley:2002gn,Barr:1993rx,Abouzaid:2010ny,Gibbons:1993zq}
[$\eta_{_{ab}}\equiv A(K_L\to\pi^a\pi^b)/ A(K_S\to\pi^a\pi^b)$]
\be\label{eq:exp}
{\rm Re} \left(\varepsilon'/\varepsilon\right)\; =\;
\frac{1}{3} \left( 1   -\left|\frac{\eta_{_{00}}}{\eta_{_{+-}}}\right|\right) \; =\;
(16.8 \pm 2.0) \times  10^{-4}
\ee
demonstrates the existence of direct CP violation in the $K\to 2\pi$ decay amplitudes.
When CP violation is turned on, the amplitudes $A_I$
acquire imaginary parts. To first order in CP violation,
\begin{equation}
\varepsilon'\;  = - \frac{i}{\sqrt{2}} \, e^{i ( \chi_2 - \chi_0 )} \,
\frac{\mathrm{Re} A_{2}}{ \;\mathrm{Re} A_{0}} \,
\left[
\frac{\mathrm{Im} A_{0}}{ \mathrm{Re} A_{0}} \, - \,
\frac{\mathrm{Im} A_{2}}{ \mathrm{Re} A_{2}} \right] ,
\label{eq:cp1}
\end{equation}
where the strong phases $\chi_I$ can be identified with the S-wave $\pi\pi$
scattering phase shifts at $\sqrt{s}=m_K$, up to isospin-breaking effects \cite{Cirigliano:2009rr,Cirigliano:2003gt}.
The phase $\phi_{\varepsilon'} =\chi_2 - \chi_0 + \pi/2 = (42.5\pm 0.9)^\circ$
is very close to the so-called superweak phase,
$\phi_\varepsilon \approx \tan^{-1}{
\left[
2 (m_{K_L}-m_{K_S})/(\Gamma_{K_S}-\Gamma_{K_L})
\right]}
= (43.52\pm 0.05)^\circ$,
implying that $\cos{(\phi_{\varepsilon'} -\phi_\varepsilon)}\approx 1$.
The CP-conserving amplitudes $\mathrm{Re} A_{I}$ can be set to their experimentally determined values,
avoiding in this way the large uncertainties associated with the
hadronic matrix elements of the four-quark operators in $\cL_{\mathrm{eff}}^{\Delta S=1}$.
Thus, one only needs a first-principle calculation of the CP-odd amplitudes $\mathrm{Im} A_{0}$
and $\mathrm{Im} A_{2}$; the first one is completely dominated by the strong penguin operator
$Q_6$, while the leading contribution to the second one comes from the
electromagnetic penguin $Q_8$.
Fortunately, those are precisely the only operators that are
well approximated through a large-$N_C$ estimate of LECs, because their anomalous dimensions
are leading in $1/N_C$.
Owing to the large ratio  $\mathrm{Re} A_{0}/ \mathrm{Re} A_{2}$, isospin violation
plays also an important role in $\varepsilon'/\varepsilon$ \cite{Cirigliano:2003gt}.
The one-loop $\chi$PT enhancement of the isoscalar amplitude \cite{Pallante:1999qf,Pallante:2001he} 
destroys an accidental LO cancellation of the two terms in \eqn{eq:cp1} \cite{Buchalla:1995vs,Bertolini:1998vd,Hambye:1999yy},
bringing the SM prediction
of $\varepsilon'/\varepsilon$ in good agreement with the experimental measurement in Eq.~(\ref{eq:exp})
\cite{Cirigliano:2011ny,Pallante:1999qf,Pallante:2001he}:
\be\label{eq:finalRes}
\mbox{Re}\left(\epsilon'/\epsilon\right) \; =\;
\left(19\pm 2_{\mu}\, {{}_{-6}^{+9}}_{m_s} \pm 6_{1/N_C}\right) \times 10^{-4}\, .
\ee

\section{Rare and Radiative Decays}

Kaon decays mediated by flavour-changing neutral currents are suppressed in the SM and
their main interest, other than their own understanding, relies on the possible observation of new physics effects.  Most of these processes ($K\to\gamma^{(*)}\gamma^{(*)}$, $K\to\ell^+\ell^-$, $K\to\pi\gamma^{(*)}\gamma^{(*)}$, $K\to\pi\ell^+\ell^-$\ldots)
are dominated by long-distance contributions which can be analyzed with $\chi$PT techniques
\cite{D'Ambrosio:1986ze,Goity:1986sr,Ecker:1987fm,Ecker:1987hd,Cappiello:1988yg,Ecker:1991ru,Cohen:1993ta,Kambor:1993tv,Cappiello:1992kk,D'Ambrosio:1996zx,D'Ambrosio:1998yj,Donoghue:1994yt,Ecker:1993cq,Bijnens:1992ky}.
However, there are also processes governed by short-distance amplitudes, such as $K \rightarrow\pi \nu \bar{\nu}$ \cite{Buchalla:1995vs,Buras:2005gr,Brod:2010hi}.
The decays $K_L\to\pi^0 e^+e^-$ and $K_L\to\pi^0\nu\bar\nu$ provide very interesting measures of CP violation. The first one receives contributions from both direct and indirect CP violation which dominate the decay amplitude; the CP-conserving contribution being suppressed by an additional power of $\alpha$ \cite{Ecker:1987hd}. The decay $K_L\to\pi^0\nu\bar\nu$ violates CP and it is completely dominated by the direct-CP contribution.
A detailed overview of rare K decays with a more complete reference list can be found in Refs.~\cite{Cirigliano:2011ny,Pich:2012vd}.

\section*{Acknowledgments}

Work supported in part by the Spanish Government~[grants FPA2011-23778 and CSD2007-00042~(Consolider Project CPAN)] and the Generalitat Valenciana [PrometeoII/2013/007].

\end{document}